\documentclass[prl,reprint]{revtex4-1}

\usepackage{graphicx}
\usepackage{bm}
\usepackage{hyperref}
\usepackage{color}

\begin{document}

\title{Morphing of 2D Hole Systems at $\nu=3/2$ in Parallel Magnetic Fields: Compressible, Stripe, and Fractional Quantum Hall Phases}

\author{Yang Liu}
\affiliation{Department of Electrical Engineering,
Princeton University, Princeton, New Jersey 08544}
\author{ M. A. Mueed}
\affiliation{Department of Electrical Engineering,
Princeton University, Princeton, New Jersey 08544}
\author{ Md. Shafayat Hossain}
\affiliation{Department of Electrical Engineering,
Princeton University, Princeton, New Jersey 08544}
\author{S. Hasdermir}
\affiliation{Department of Electrical Engineering,
Princeton University, Princeton, New Jersey 08544}
\author{L.N.\ Pfeiffer}
\affiliation{Department of Electrical Engineering,
Princeton University, Princeton, New Jersey 08544}
\author{K.W.\ West}
\affiliation{Department of Electrical Engineering,
Princeton University, Princeton, New Jersey 08544}
\author{K.W.\ Baldwin}
\affiliation{Department of Electrical Engineering,
Princeton University, Princeton, New Jersey 08544}
\author{M.\ Shayegan}
\affiliation{Department of Electrical Engineering,
Princeton University, Princeton, New Jersey 08544}

\date{\today}

\begin{abstract}
A transport study of two-dimensional (2D) holes confined to wide GaAs quantum wells provides a glimpse of a subtle competition between different many-body phases at Landau level filling $\nu=3/2$ in tilted magnetic fields. At large tilt angles ($\theta$), an anisotropic, stripe (or nematic) phase replaces the isotropic compressible Fermi sea at $\nu=3/2$ if the quantum well has a symmetric charge distribution. When the charge distribution is made asymmetric, instead of the stripe phase, an even-denominator fractional quantum state appears at $\nu=3/2$ in a range of large $\theta$, and reverts back to a compressible state at even higher $\theta$. We attribute this remarkable evolution to the significant mixing of the excited and ground-state Landau levels of 2D hole systems in tilted fields.
\end{abstract}


\maketitle

A strong magnetic field perpendicular to a 2D electron system (2DES) quantizes the electron kinetic energy into a set of highly-degenerate Landau levels (LLs). The dominating Coulomb interaction then gives rise to numerous, exotic quantum many-body phases \cite{Shayegan.Flatland.2006,
  Jain.CF.2007}. When the Fermi energy ($E_F$) lies in an $N=0$ LL, there is a compressible Fermi sea of composite fermions at LL filling factors $\nu=1/2$ and 3/2 while numerous fractional quantum Hall states (FQHSs) are observed at nearby odd-denominator $\nu$ \cite{Shayegan.Flatland.2006,
  Jain.CF.2007, Tsui.PRL.1982, Willett.PRL.1993, Halperin.PRB.1993}. In $N\geq 2$ LLs, FQHSs are typically absent and anisotropic phases dominate at half-filled LLs, e.g., at $\nu=9/2$ and 11/2 as the system breaks the rotational symmetry and forms unidirectional charge density waves -- the so-called stripe (or nematic) phases \cite{Lilly.PRL.1999, Du.SSC.1999, Fradkin.ARCMP.2010}. The intermediate $N=1$ LL is special. The electrons exhibit FQHSs not only at odd-denominator $\nu$ but also at the even-denominator fillings $\nu=5/2$ and 7/2 \cite{Shayegan.Flatland.2006,
  Jain.CF.2007, Willett.PRL.1987}. The latter are believed to be the Moore-Read Pfaffian state \cite{Moore.Nuc.Phy.1991}, obey non-Abelian statistics, and be of potential use in topological quantum computing \cite{Nayak.Rev.Mod.Phys.2008}. The application of parallel magnetic field ($B_{||}$) or pressure can break the rotational symmetry and introduce LL mixing, leading to the destruction of the $\nu=5/2$ FQHS and stabilization of the stripe phase in the $N=1$ LL \cite{Pan.PRL.1999.Stripe, Lilly.PRL.83.1999, Samkharadze.NatPhys.2015, Xia.PRL.2010, Liu.PRB.2013B}.

In GaAs two-dimensional hole systems (2DHSs), the spin-orbit coupling mixes harmonic oscillators with different Landau and spin indices and leads to a complex set of LLs \cite{Winkler.SOC.2003}. Nevertheless, in narrow quantum wells (QWs), the 2DHS is compressible at $\nu=1/2$ and 3/2 and numerous odd-denominator FQHSs are still prevalent as the filling deviates from $\nu=1/2$ and 3/2, qualitatively similar to those in 2DESs. However, the even-denominator FQHSs at $\nu=5/2$ and 7/2 are very weak \cite{Manoharan.PRB.1994, Shayegan.PhysicaE.2000}, and instead stripe phases are typically observed at these fillings, particularly at low densities \cite{Shayegan.PhysicaE.2000, Manfra.PRL.2007, Koduvayur.PRL.2011}. Here, we report transport measurements in 2DHSs confined in \emph{wide} GaAs QWs and subjected to strong $B_{||}$. We observe a remarkable metamorphosis of the ground state at $\nu=3/2$. The compressible Fermi sea seen at $\nu=3/2$ turns into a stripe phase when we apply a sufficiently large $B_{||}$ to a symmetric QW. The stripe phase can be destabilized in asymmetric QWs and, strikingly, an even-denominator FQHS forms at $\nu=3/2$ at intermediate $B_{||}$. At larger $B_{||}$, the $\nu=3/2$ FQHS disappears and the 2DHS reverts back to becoming compressible. Our results highlight the rich and subtle many-body phenomena manifested by high-quality 2DHSs.

Our samples were grown by molecular beam epitaxy, and each consists of a GaAs QW (well widths $W=35$ or 30 nm) which is bounded on either side by undoped Al$_{0.3}$Ga$_{0.7}$As spacer layers and C $\delta$-doped layers. They have as grown densities $p\simeq 1$ to $1.5\times 10^{11}$ cm$^{-2}$ and high mobility $\mu \simeq$ 100 m$^2$/Vs. Each sample has a van der Pauw geometry, with alloyed InZn contacts at the four corners of a $4\times 4$ mm$^2$ piece. We carefully control the density and the charge distribution symmetry in the QW by applying voltage biases to the back- and front-gates \cite{Suen.PRL.1994, *Liu.PRL.2014, Shabani.PRB.2013}. For the low-temperature measurements, we use a dilution refrigerator with a sample platform which can be rotated \textit{in-situ} in the magnetic field to induce a parallel field component $B_{||}$ along the $x$-direction (see Fig. 1(c)). We use $\theta$ to express the angle between the field and the normal to the sample plane, and denote the longitudinal resistances measured along and perpendicular to the direction of $B_{||}$ as $R_{xx}$ and $R_{yy}$, respectively (Fig. 1(c)).

\begin{figure}[t]
\includegraphics[width=0.48\textwidth]{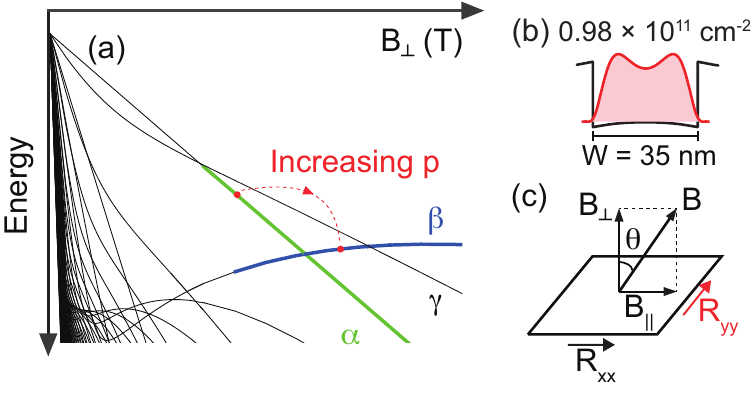}
\caption{(color online) (a) Schematic LL diagram. The light-hole-like $\beta$-level (blue) crosses the heavy-hole-like $\alpha$-level (green) as a function of $B_{\perp}$. As indicated by the red arrow, the Fermi energy $E_F$ at $\nu=3/2$ moves from the $\alpha$- to the $\beta$-level as we increase the density. (b) Calculated charge distribution and potential of the 35-nm-QW 2DHS at $p=0.98\times 10^{11}$ cm$^{-2}$. (c) Measurement setup. $B_{||}$ is applied by tilting the sample in field, and $R_{xx}$ and $R_{yy}$ denote the longitudinal resistance measured along and perpendicular to $B_{||}$.}
\end{figure}

\begin{figure}[t]
\includegraphics[width=0.48\textwidth]{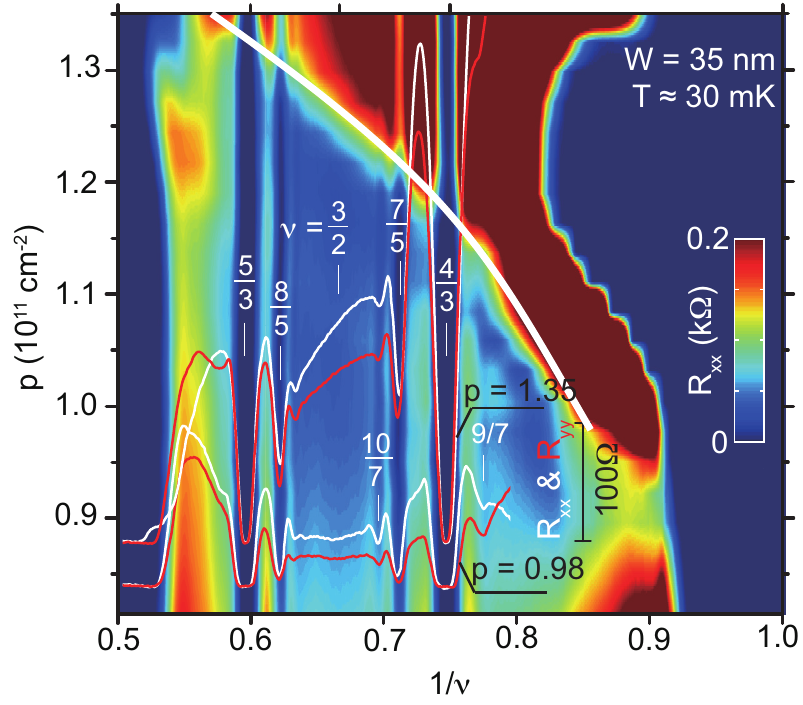}
\caption{(color online) Color-coded plot of
  ${R_{xx}}$ measured from the 35-nm-wide QW sample at
  $\theta=0^{\circ}$ and different densities. A sharp phase transition
  can be seen as a steep increase of ${R_{xx}}$ at the boundary marked
  by the solid white line, signaling a LL crossing. The $R_{xx}$
  (white) and $R_{yy}$ (red) traces are taken at $p=0.98$ and
  $1.35\times 10^{11}$ cm$^{-2}$, before and after the crossing,
  respectively. Despite the significant increase in $R_{xx}$ and $R_{yy}$, the
  2DHS in both cases is isotropic and exhibits strong FQHSs at
  $\nu=4/3$, 5/3, etc., and a compressible Fermi sea at $\nu=3/2$,
  indicating the two crossing LLs are both $N=0$.}
\end{figure}

\begin{figure*}[t]
\includegraphics[width=0.95\textwidth]{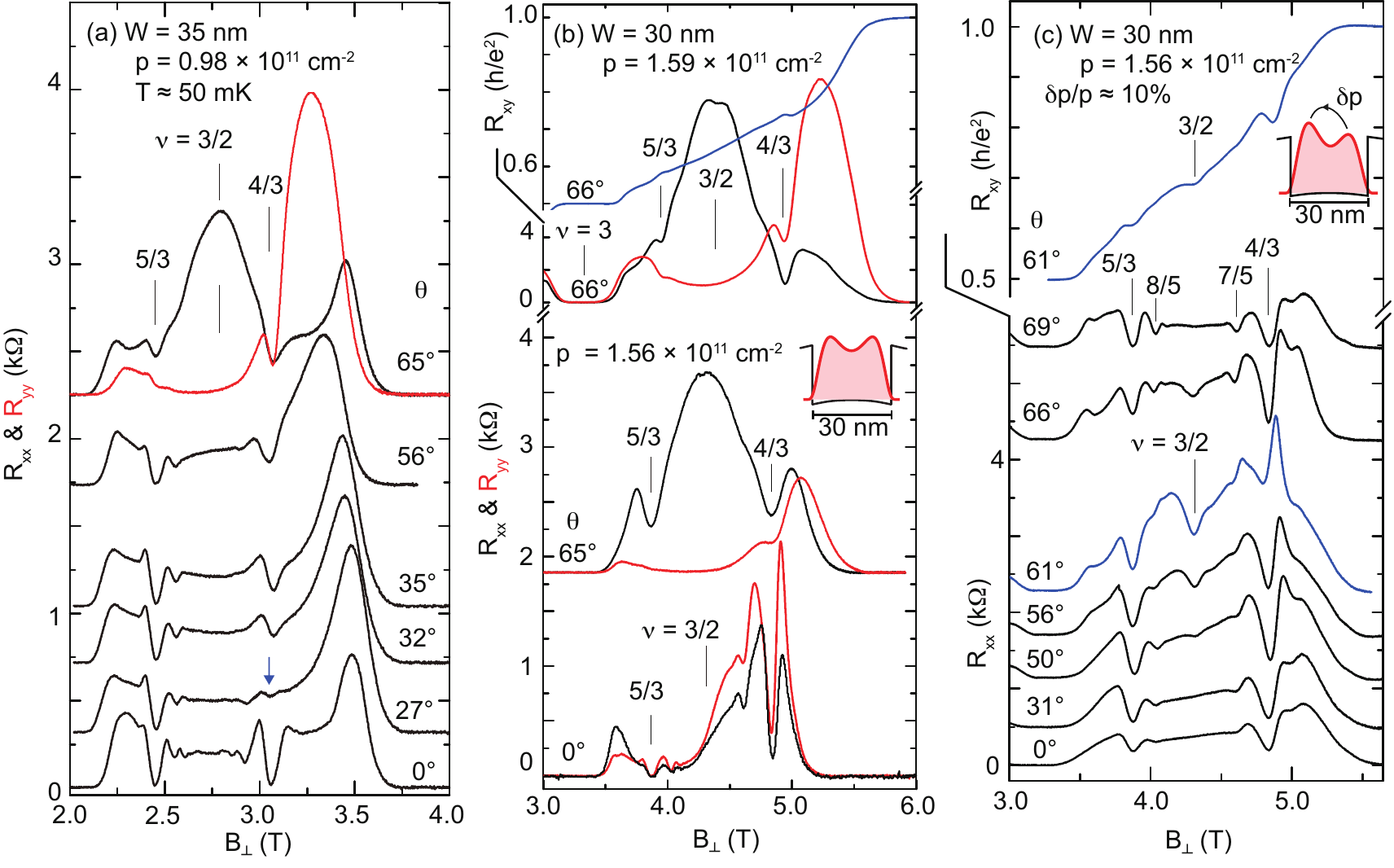}
\caption{(color online) (a) $R_{xx}$ and $R_{yy}$
  for the 35-nm-QW 2DHS at density $p\simeq 0.98\times 10^{11}$
  cm${^{-2}}$ and different tilt angles ($\theta$). At
  $\theta=65^{\circ}$, the compressible Fermi sea at $\nu=3/2$ turns into a stripe
  phase and the $R_{xx}$ minima at $\nu=4/3$ and 5/3 become
  weak. (b) Magneto-resistances for the symmetric
  30-nm-QW samples. (c) In a slightly asymmetric
  QW, no stripe phase but a Fermi sea is seen at very large
  $\theta$. Remarkably, near $\theta\simeq 61^{\circ}$, an even-denominator FQHS is seen at
  $\nu=3/2$. In all panels, the traces are shifted vertically for clarity.}
\end{figure*}

Although the main focus of our study is the state of the 2DHS near $\nu=3/2$ in \emph{tilted} magnetic fields, the data at $\theta=0$ are also very intriguing. Figure 2 shows $R_{xx}$ measured from a symmetric 35-nm-QW 2DHS at $\theta=0^{\circ}$ and different densities. Strong odd-denominator FQHSs are seen as vertical, low-resistance (blue) stripes at $\nu=5/3$, 8/5, 7/5, and 4/3. With increasing density, $R_{xx}$ steeply increases above a boundary marked by the white solid line. This sharp transition is a signature of a LL crossing near $\nu=3/2$. We indeed expect such a crossing from the typical LL diagram (see Fig. 1(a)) for our wide-QW 2DHSs \cite{Note1}. 
As depicted in Fig. 1(a), the light-hole-like $\beta$-level (blue) crosses the heavy-hole-like $\alpha$-level (green) as a function of $B_{\perp}$, and the Fermi energy $E_F$ at $\nu=3/2$ moves from $\alpha$- to $\beta$-level as we increase the density. Because of the strong mixing of the heavy-hole and light-hole energy levels through the spin-orbit interaction, the LLs for holes confined to wide GaAs QWs are nonlinear with $B_{\perp}$ and show multiple crossings \cite{Winkler.SOC.2003}. In previous studies, the crossing between the \textit{lowest} two LLs, those labeled $\beta$ and $\gamma$ in Fig. 1(a), was shown to lead to novel many-body phenomena, such as the stabilization of two-component QHSs at $\nu=1/2$ and $\nu=1$, as well as the appearance of an anisotropic Wigner crystal near $\nu=1/2$ \cite{Graninger.PRL.2011, *Liu.PRB.2015B, Liu.PRB.2014, Liu.cond.mat.2016}. Here we focus on data taken near the LL crossing at $\nu=3/2$.

In Fig. 1(a) we denote the two crossing LLs near $\nu=3/2$ by $\alpha$ and $\beta$. According to the calculations, the $\alpha$-level is a heavy-hole LL, and has a pure orbital ($N=0$) and spin ($J_z=-3/2$) character. It does not mix with other LLs at $\theta=0$. The $\beta$-level, on the other hand, has a light-hole character, and has mixing from the higher LLs. At low densities and small magnetic fields, i.e., below the boundary marked by the white line in Fig. 2, $E_F$ at and near $\nu=3/2$ is in the $\alpha$-level. When $p$ is increased, $E_F$ moves to the $\beta$-level above the boundary. The FQHSs at $\nu=5/3$ and 4/3 are seen on both sides of the crossing. They do not disappear at the crossing, although are clearly weaker above the white boundary. This is qualitatively similar to what is seen at the crossing of two parallel-spin LLs, and is in contrast to the crossing of two opposite-spin LLs where the FQHSs become very weak or disappear \cite{Liu.PRL.2011, Liu.PRB.2011, *Liu.PRL.2011B, *Liu.PRB.2015}. We conclude that the two crossing LLs at $\nu=3/2$ in Fig. 2 have similar spin (or pseudo-spin) orientations. Also our observation of well-developed $\nu=7/5$ and 8/5 FQHSs on both sides of the white boundary (see $R_{xx}$ and $R_{yy}$ traces at $p\simeq 0.98$ and $1.35\times 10^{11}$ cm${^{-2}}$ included in Fig. 2) implies that the two crossing levels have a primarily $N=0$ character. Note that the equivalent FQHSs in an $N=1$ LL (e.g., those at $\nu=12/5$ and 13/5 in GaAs 2D electrons) are extremely fragile and are only seen in the highest quality samples and at extremely low temperatures \cite{Xia.PRL.2004, Samkharadze.PRB.2011B}. While the $\theta=0$ data of Fig. 2 strongly suggest that the two crossing levels near $\nu=3/2$ have an $N=0$ character, we emphasize that at high $B_{||}$ where we observe the new phases of the 2DHS at $\nu=3/2$, the 2DHS LLs are very likely to be mixed with higher LLs.

We now concentrate on data taken in tilted fields. Figure 3(a) shows data from the 35-nm-wide QW at $p\simeq 0.98\times 10^{11}$ cm${^{-2}}$ and for different tilt angles $\theta$ \cite{Note2}. 
The traces presented in this figure were taken at $T\simeq 50$ mK, somewhat higher than in Fig. 2 data, because of the higher base temperature of the dilution refrigerator with the rotatable platform. The large-$\theta$ traces in Fig. 3(a) highlight our first key observation: the $B_{||}$-induced transition from a compressible to a stripe phase at $\nu=3/2$ \cite{Note3}.
The $\nu=3/2$ compressible state and the surrounding odd-denominator FQHSs remain stable as we increase $\theta$ to $\sim 56^{\circ}$. Data at $\theta\simeq 65^{\circ}$, however, reveal a weakening of the FQHSs and a very strong transport anisotropy: $R_{xx}$ at $\nu=3/2$ shows a peak that is about two orders of magnitude larger than $R_{yy}$ which exhibits a minimum. These features have a striking resemblance to what is seen when stripe phases form in GaAs 2DESs at large $B_{||}$ in $N\geq 1$ LLs ($\nu\geq 5/2$) \cite{Pan.PRL.1999.Stripe, Lilly.PRL.83.1999}. Note that the ``hard-axis'' in Fig. 3(a) data is along the $B_{||}$, also similar to the 2DES data in previous experiments \cite{Pan.PRL.1999.Stripe, Lilly.PRL.83.1999}. The surprise is that in our 2DHSs samples we observe stripe-phase features at $\nu=3/2$, and when the relevant LLs clearly have an $N=0$ character at $\theta=0^{\circ}$. We conclude that $B_{||}$ induces a severe mixing of the higher-index LLs, leading to the stripe phase at fillings as small as $\nu=3/2$.

In Fig. 3(b), we show data for a 30-nm-QW sample with $p\simeq 1.56\times 10^{11}$ cm$^{-2}$ at $\theta=0^{\circ}$ and $65^{\circ}$; data at $\theta=66^{\circ}$ are taken from another sample with a slightly higher density $p\simeq 1.59\times 10^{11}$ cm$^{-2}$. The charge distributions in these samples are symmetric and the 2DHSs in both samples show a strong FQHS at $\nu=1/2$ when $\theta=0^{\circ}$. In both samples, the presence of an anisotropic phase at $\nu=3/2$ at large $\theta$ is clear. The similarity of the $\theta=66^{\circ}$ traces to those at $\theta=65^{\circ}$ for the 35-nm-sample (Fig. 3(a)) is remarkable. The $\theta=65^{\circ}$ traces in Fig. 3(b) also clearly show an anisotropic phase at $\nu=3/2$, but for $B_{\perp} \gtrsim 5$ T, transport is essentially isotropic. We do not know the origin of the different behavior for the two samples \cite{Note4}.
Regardless of the difference, it is clear that in all three samples of Figs. 3(a) and 3(b), $R_{xx} \gg R_{yy}$ at $\nu=3/2$ at large $\theta$. The data provide strong evidence for a transition from an essentially isotropic, compressible phase at $\nu=3/2$ at $\theta=0$ to an anisotropic phase at large tilt angles. Note also that in all these three samples, at $\theta \simeq 66^{\circ}$ the longitudinal resistances exhibit minima and the Hall resistance exhibits developing plateaus at $\nu=4/3$. The coexistence of a stripe phase and a FQHS at a nearby filling is noteworthy. Our data (not shown) also indicate that the anisotropic phase only appears at very low temperatures; when we raise $T$ to $\simeq 200$ mK, $R_{xx}$ and $R_{yy}$ become nearly isotropic. This is similar to what is observed for $B_{||}$-induced anisotropic phases, e.g., at $\nu=5/2$, in GaAs 2D electrons \cite{Pan.PRL.1999.Stripe, Lilly.PRL.83.1999, Xia.PRL.2010, Liu.PRB.2013B}.

Data presented in Fig. 3(c) reveal yet another twist in the fate of the ground-state at $\nu=3/2$ in 2DHSs! Here we show data for a third 30-nm-QW sample. It is from the same wafer as the Fig. 3(b) data and has a very similar density ($p\simeq 1.56\times 10^{11}$ cm$^{-2}$) but the charge distribution is slightly asymmetric (about 10\%, see inset in Fig. 3(c)) as judged from the absence of a FQHS at $\nu=1/2$ at $\theta=0^{\circ}$ \cite{Note5}.
Surprisingly, when tilted to high $\theta$, there is no anisotropic phase; instead, an even-denominator FQHS develops at $\nu=3/2$. The state is strongest at $\theta=61^{\circ}$ with a deep $R_{xx}$ minimum accompanied by a clear $R_{xy}$ plateau at $(2/3)(h/e^2)$. At higher $\theta$ the $\nu=3/2$ FQHS becomes weaker and completely disappears at $\theta=69^{\circ}$ where it is replaced by a compressible state. The $\theta=69^{\circ}$ trace is by itself intriguing as the FQHSs $\nu=4/3$ and 5/3 are quite strong, even stronger than they are at $\theta=0^{\circ}$, and there are also developing higher order FQHSs at $\nu=7/5$ and 8/5.

The unusual evolution we observe for the ground-state as a function of $\theta$ and charge distribution symmetry attests to the subtle and complex nature of 2DHS LLs in the presence of both perpendicular and parallel magnetic fields. Unfortunately, calculations of the LLs in multi-subband 2DHSs at finite $\theta$ are very challenging and unavailable; indeed, we hope that our results would provide stimulus for such calculations. We note that in GaAs 2DESs, no stripe phases are observed, or expected, when $E_F$ lies in an $N=0$ LL. Our data therefore suggest that there is a significant mixing of the higher-index LLs at large $B_{||}$ in 2DHSs confined to wide QWs, and the mixing leads to a close competition between various many-body states, including the stripe phase and the even-denominator FQHS. Given the complex hole band-structure and LLs, such mixing is certainly plausible. 

The FQHS we observe at $\nu=3/2$ is particularly intriguing. In very high quality, single-layer 2D systems, even-denominator FQHSs are seen at $\nu=5/2$ (and 7/2) when $E_F$ lies in an $N=1$ LL \cite{Willett.PRL.1987, Manoharan.PRB.1994, Falson.Nat.Phys.2015, Note6}.
The origin of these FQHSs is not yet fully established. They might be a fully spin-polarized, one-component, Moore-Read Pfaffian state \cite{Moore.Nuc.Phy.1991} and obey non-Abelian statistics \cite{Nayak.Rev.Mod.Phys.2008}, or be the un-polarized, two-component ($\Psi_{331}$) state \cite{Halperin.HPA.1983}. In 2D systems with a layer or subband degree of freedom, even-denominator FQHSs are also seen at $\nu=1/2$ (and 3/2) when the $N=0$ LLs from different layers or subbands are close in energy \cite{Suen.PRL.1992, Eisenstein.PRL.1992, Suen.PRL.1994, Shabani.PRB.2013, Liu.PRL.2014, Liu.PRB.2014}. These are generally believed to be two-component FQHSs. 

Compared to the even-denominator FQHSs described above, the $\nu=3/2$ FQHS we observe near $\theta\simeq 61^{\circ}$ in Fig. 3(c) has some unique characteristics. For example, the two-component $\nu=1/2$ state seen in wide QWs is strongest when the charge distribution is symmetric, and turns into a compressible state for sufficient ($\sim$10\%) charge distribution asymmetry \cite{Suen.PRL.1994, Shabani.PRB.2013, Liu.PRL.2014}. This is opposite to what we see in Figs. 3(b) and (c) where the charge distribution asymmetry \textit{helps} to stabilize the 3/2 FQHS. As another example, we note that the 5/2 FQHS in GaAs 2DESs is observed at $\theta=0$ but becomes unstable at large $B_{||}$ and turns into a stripe phase. Again, this is in contrast to the behavior in Fig. 3(c). We suggest that it is the mixing of $N=1$ and $N=0$ LLs that leads to the stabilization of the 3/2 FQHS at intermediate $\theta$ in Fig. 3(c). There is indeed experimental evidence \cite{Liu.PRL.2011}, backed by theoretical calculations \cite{Papic.PRL.2012}, that the $\nu=5/2$ FQHS, which is formed in an $N=1$ LL, becomes more robust when there is a nearby $N=0$ LL. 

In conclusion, our results for 2DHSs confined in wide GaAs QWs reveal a remarkable evolution of different many-body phases at $\nu=3/2$ in tilted magnetic fields: a compressible state, an anisotropic (stripe) phase, and an incompressible FQHS. We attribute this metamorphosis to the subtle nature of the mixed 2DHS LLs in tilted fields. The detailed character of these LLs and how they affect the ground states of the 2DHS await future theoretical and experimental studies.

\begin{acknowledgments}
We acknowledge support by the DOE BES (DE-FG02-00-ER45841) grant for measurements, and the NSF (Grants DMR-1305691, and MRSEC DMR-1420541), the Gordon and Betty Moore Foundation (Grant GBMF4420), and Keck Foundation for sample fabrication and characterization. We thank R. Winkler for providing the calculated Landau level fan diagram in Fig. 1(a) and the charge distribution and potentials shown in Figs. 1(b) and 3(b). Our measurements were partly performed at the National High Magnetic Field Laboratory (NHMFL), which is supported by the NSF Cooperative Agreement DMR-1157490, by the State of Florida, and by the DOE. We thank T. Murphy, J. H. Park and G. E. Jones at NHMFL for technical assistance.
\end{acknowledgments}

\bibliographystyle{apsrev4-1} 
\bibliography{../bib_full}

\end{document}